\documentclass[11pt]{amsart}
\usepackage{amsmath,amssymb,amsthm} 
\usepackage[T1]{fontenc}
\usepackage[utf8]{inputenc}
\usepackage{lmodern}
\usepackage{graphicx}
\usepackage{subcaption}
\usepackage{chronology}
\usepackage{csquotes}
\usepackage[english]{babel}
\usepackage{url}
\usepackage{amsthm}
\usepackage{amsmath}
\usepackage{amsfonts}
\usepackage{amssymb}
\usepackage[shortlabels]{enumitem}
\usepackage{listings}
\usepackage{xcolor}
\usepackage{bbm}
\usepackage{rotating}
\usepackage{tikz}
\usepackage{placeins}
\usepackage{amsaddr}
\usepackage[round, compress]{natbib}
\usepackage{hyperref}
\usepackage{comment}
\usepackage[margin=1.25in]{geometry}

\numberwithin{equation}{section}

\newcommand{\E}{\mathbb{E}}

\newcommand*{\affmark}[1][*]{\textsuperscript{#1}}

\newcommand{\subtitle}[1]{%
	\posttitle{%
		\par\end{center}
	\begin{center}\large#1\end{center}
	\vskip0.1em}}%

\begin{document}

\title[The cost of misspecifying price impact]{The cost of misspecifying price impact}

\author{%
Natascha Hey\affmark[1], Jean-Philippe Bouchaud\affmark[1,]\affmark[2,]\affmark[3],\\ Iacopo Mastromatteo\affmark[2], Johannes Muhle-Karbe\affmark[4],\\ Kevin Webster\affmark[4]}
\address{\affmark[1]Chair of Econophysics and Complex Systems, \'Ecole Polytechnique
\\
\affmark[2]Capital Fund Management
\\
\affmark[3]Acad\'emie des Sciences
\\
\affmark[4]Department of Mathematics, Imperial College London 
}
\email{natascha.hey@ladhyx.polytechnique.fr}
\email{jean-philippe.bouchaud@cfm.com}
\email{iacopo.mastromatteo@cfm.com}
\email{jmuhleka@ic.ac.uk}
\email{kwebster@ic.ac.uk}

\date{\today}

\maketitle

\begin{abstract}
Portfolio managers' orders trade off return and trading cost predictions. Return predictions rely on alpha models, whereas price impact models quantify trading costs. This paper studies what happens when trades are based on an incorrect price impact model, so that the portfolio either over- or under-trades its alpha signal. We derive tractable formulas for these misspecification costs and illustrate them on proprietary trading data.  The misspecification costs are naturally asymmetric: underestimating impact concavity or impact decay shrinks profits, but overestimating concavity or impact decay can even turn profits into losses.
\end{abstract}

\section{Introduction}

Price impact refers to price movements induced purely by trading flows, independently of their information content. For large investors, such adverse price moves caused by their own trades are the main source of transaction costs.\footnote{In this regime, other sources of costs such as proportional bid-ask spreads are of secondary concern and hence will be disregarded in the following; the interested reader is referred to~\cite{martin2011mean} for a detailed  discussion of the role of such linear costs for the design of trading strategies that operate on a smaller scale.} Price impact models are thus essential tools in algorithmic trading, allowing investment teams to estimate the effect of their trades on asset prices and thereby design, size and deploy systematic strategies. For instance, (capacity constrained) statistical arbitrage portfolios seek to achieve the best trade-off between price predictions and trading costs.\footnote{If transaction costs do no constrain the size of the portfolio tightly enough, it may also be risk constrained, cf., e.g., \cite{Garleanu2013} and the references therein.
}

The price forecast is commonly called an \emph{alpha signal}. Turning the latter into optimal trades in turn requires an appropriate price impact model. Many empirical studies find a concave \emph{nonlinear} relationship between order size and the price impact of sizeable trades, see, e.g.,~\cite*{BouchaudBook,webster.23} and the references therein for an overview. More specifically, two essential parameters emerge when determining the evolution of price impact: \emph{concavity} and \emph{impact decay}. 

\emph{Concavity} describes how the marginal build-up of price impact slows beyond a certain order size: larger orders are cheaper than a linear model suggests. \emph{Impact decay} describes how long trade-induced price moves linger in the market and affects how quickly portfolios can turn over.  

This paper quantifies how incorrect calibration of these price impact parameters leads to significant misspecification costs for traders and portfolio managers. This allows to go beyond purely statistical parameter estimates by establishing P\&L-driven error bounds for price impact parameters, which are crucial to any team estimating the actual trading capacity of their alpha signals.

To carry out this analysis, we derive closed-form expressions for misspecification costs in a nonlinear but nevertheless tractable price impact model introduced by~\cite*{Alfonsi2010}. This model was originally proposed and studied for optimal execution problems, but in fact admits explicit formulas for optimal trades with general alpha signals. 

These in turn allow one to quantify how wrong model parameters can be before turning a profitable trading strategy for a given alpha signal into an unprofitable one. For example, linear price impact models as in~\cite{Obizhaeva2013} overestimate the transaction costs of large trades and thereby imply overly conservative trading strategies. However, they rarely turn a winning strategy into a losing one. Conversely, even relatively small overestimates of impact concavity can lead to overly aggressive trading that leads to losses despite an accurate alpha signal. Similarly, underestimating impact decay leads to conservative strategies that trade slower than optimal, which sacrifices some trading opportunities but does not lead to dramatic losses. In contrast, overly aggressive trades based on overestimates of impact decay can quickly turn a portfolio's P\&L negative.

\begin{figure}[ht]
\captionsetup{width=1.0\linewidth}
    \includegraphics[width=0.9\linewidth]{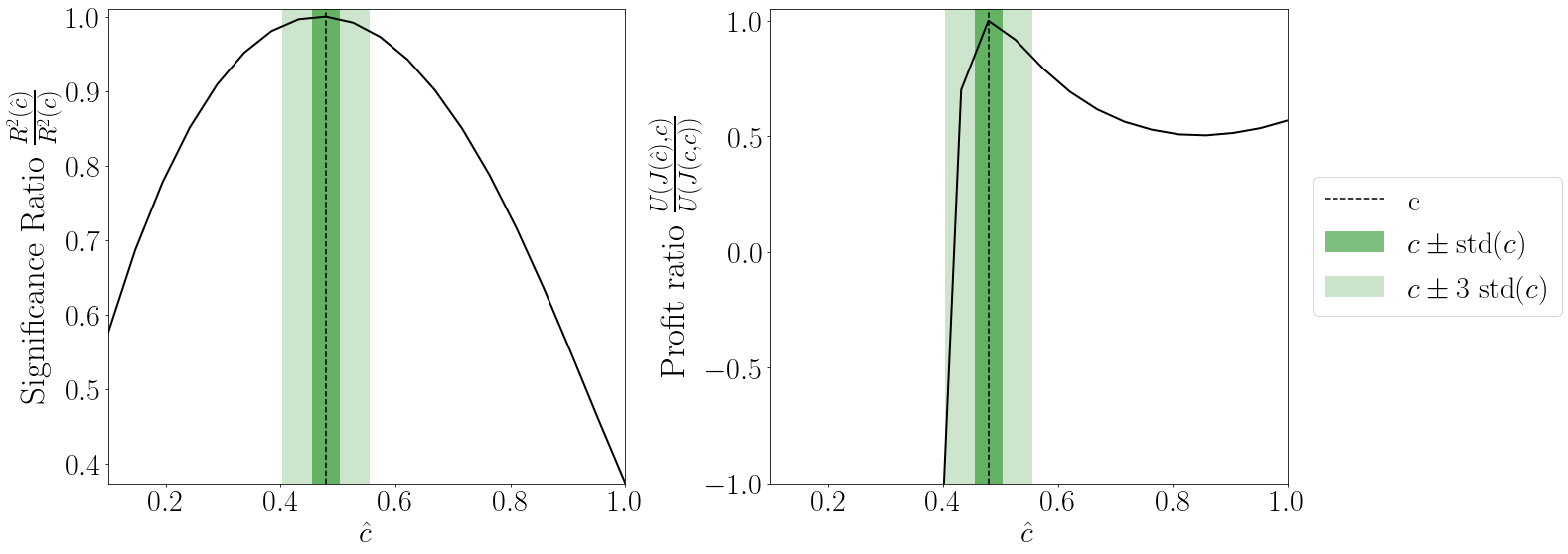}
	\caption{Left panel: Symmetric statistical significance ratio $R^2(\hat c)/R^2(c)$ for point estimate $c=0.48$ plotted against misspecified impact concavity $\hat c$. Right panel: Asymmetric profit ratio $U(\hat{c})/U(c)$ for point estimate $c=0.48$ plotted against misspecified impact concavity $\hat c$. The dark and light shaded areas are bootstrap confidence intervals around $c$.}
    \label{fig:intro}
\end{figure}

These fundamental  asymmetries are illustrated in Figure~\ref{fig:intro}, which contrasts the symmetric shape of the statistical model fit with the highly asymmetric nature of the corresponding P\&Ls.

The remainder of the paper is organized in three parts:
\begin{itemize}
\item \emph{Optimal Trading Strategy:} Section~\ref{sec:optimal} derives closed-form formulas for trading general alpha signals with non-linear price impact. These can be implemented into automated trading algorithms or used in Transaction Cost Analysis (TCA).
\item \emph{Empirical Estimation:} Section~\ref{sec:fitting} estimates the impact decay and concavity parameters of the non-linear price impact model using proprietary trading data. 
\item \emph{Sensitivity Analysis:} Section \ref{sec:sensitivity} combines the results from the previous two sections. Using the explicit trading strategies from Section~\ref{sec:optimal}, we evaluate the P\&L implications of the parameter estimates from Section~\ref{sec:fitting}. 
\end{itemize}
Both optimal trading and parameter estimation in models with nonlinear price impact are studied in greater detail in the technical companion paper of the present study~\citep*{bouchaud.al.23}, to which we also refer for the derivations of the results presented here.

\section{Optimal Trading Strategy}\label{sec:optimal}

This section takes a price impact model as given and focuses on deriving the trading strategy that maximizes an alpha signal's value net of price impact costs. We write $Q_t$ for a portfolio's position at time $t$. Therefore, $dQ_t$ represents the trade at time $t$. Let $S$ be an It\^o process describing the midprice in the absence of trading, also called the ``unperturbed'' or ``fundamental'' price. The observed midprice at time $t$ is 
$$P_t = S_t + I_t(Q),$$
where $I_t(Q)$ is the impact caused by the trades $(Q_s)_{s \leq t}$ up to time $t$.

\subsection{The AFS model}

We focus on the price impact model introduced by~\cite{Alfonsi2010} (henceforth AFS) and studied further by~\cite{gatheral.al.12}, which captures the nonlinear nature of price impact but nevertheless remains remarkably tractable. 

In the AFS model, price impact is a nonlinear function of a moving average of current and past order flow:\footnote{To simplify the exposition, this paper focuses on price impact functions of power form and leaves the general treatment of the AFS model to the technical companion paper~\citep{bouchaud.al.23}. One can also incorporate time-dependent and even stochastic liquidity by modelling $\tau,\lambda$ as processes instead of constants. Such a model still leads to closed-form formulas, reported in the companion paper.}
\begin{equation}\label{eq:impact}
I_t = \lambda\ \text{sign}(J_t) |J_t|^{c},
\end{equation}
where $J_t$ is the exponentially weighted moving average
\begin{equation}\label{eq:volimpact}
dJ_t = -\frac{1}{\tau} J_t dt + dQ_t, \quad J_0=0.
\end{equation}
Here, $\lambda>0$ describes the magnitude of price impact, $c \in (0,1]$ measures its concavity, and $\tau>0$ describes the time scale over which impact decays.\footnote{Note that  this specification implies that impact $I_t$ relaxes all the way to zero over long time horizons, i.e., trades have no \emph{permanent} impact. But see \cite{gabaix2021search,bouchaud2022inelastic} for a detailed discussion of this point.}

For $c=1$, price impact is linear and one recovers the model of~\cite{Obizhaeva2013}. $c = 0.5$ corresponds to the ``square-root law'' practitioners commonly use in Transaction Cost Analysis (TCA). 

\subsection{The optimization problem}

Given a model for the unperturbed price $S$, a trading strategy $Q$ and an impact model $I(Q)$, a portfolio's P\&L $Y$ has the dynamics\footnote{This representation holds exactly if the trading strategy is smooth. If the trading strategy includes larger transactions such as bulk trades or diffusive purchases and salers, then there are some additional correction terms detailed in the technical companion paper.}
\begin{equation*}
dY_t = Q_t dS_t - I_t dQ_t.
\end{equation*}
For many statistical arbitrage strategies, price impact rather than risk is the main capacity constraint. In this regime, one maximizes the expected P\&L, given future return predictions. Such predictions take the form of an alpha signal:
\begin{equation*}
\alpha_t = \E_t \left[S_{t+h} - S_t\right] \quad \mbox{for some prediction horizon $h>T$.}
\end{equation*}

In addition to the \emph{level} $\alpha_t$ of the current alpha prediction, another crucial statistic in this context is its \emph{decay}, captured by its drift rate $\mu^{\alpha}_t$. For smooth alphas, this is simply the derivative $\mu^\alpha_t = \alpha'_t$.

For a given alpha signal, a risk-neutral statistical arbitrage portfolio maximizes
\begin{equation}\label{eq:goal}
\E\left[Y_T\right] = \E\left[\int_0^T (\alpha_t - I_t)dQ_t\right].
\end{equation}
That is, each trade captures the expected alpha but pays price impact since -- as noted above -- the latter does eventually disappear. 

\subsection{Solving the problem in impact space}

Maybe surprisingly, the statistical arbitrage problem~\eqref{eq:goal} has a straightforward, closed-form solution for arbitrary alpha signals, even for non-linear price impact described by the AFS model. The result hinges on a simple observation, proven in the companion paper using a technique introduced by~\cite*{Fruth2013, Fruth2019}. The key insight is that, for any choice of position $Q$ or, \emph{equivalently}, the corresponding impact variable $J$ (respectively $I$), one has
\begin{equation*}
\E\left[Y_T\right] = \E\left[\frac{1}{\tau} \int_0^T (\alpha_t J_t - \lambda |J_t|^{1+c}) dt - \int_0^T J_t d\alpha_t + \alpha_T J_T- \frac{1}{1+c}|J_T|^{1+c}\right].
\end{equation*}
Whence, by switching the control variables from positions $Q$ to the corresponding impact $J$ (``passing to \emph{impact space}''), the complex control problem~\eqref{eq:goal} becomes a simple pointwise maximization. The optimal impact in turn is\footnote{The companion paper extends this result to general non-linear price impact functions as well as stochastic liquidity.}
\begin{equation}
I^*_t = \lambda \left(J^*_t\right)^c = \begin{cases} \frac{1}{1+c}\left(\alpha_t - \tau \mu^\alpha_t\right), &t \in (0,T)\\  \alpha_T, & t=T. \end{cases} \label{eq:optim}
\end{equation}
The corresponding optimal positions can be recovered via
\begin{equation}\label{eq:portfolio}
Q^*_t = J^*_t + \frac{1}{\tau} \int_0^t J^*_s ds.
\end{equation}
Whence, even though the optimal trading strategy $Q^*$ can become rather complex, switching to impact space allows one to retain a straightforward linear relationship between the optimal policy as well as the current alpha signal and its decay. Up to a bulk trade at the terminal time that exhausts all remaining alpha, it is optimal to keep impact equal to a constant fraction of the alpha signal, adjusted for alpha decay relative to impact decay. The intuition for this adjustment is that one has to trade more aggressively and accept higher impact costs for signals that decay quickly compared to impact, as already highlighted by \cite{Garleanu2013} for linear impact models:
\begin{displayquote}
``The alpha decay is important because it determines how long time the investor can enjoy high expected returns and, therefore, affects the trade-off between returns and transactions costs.''
\end{displayquote}
The fraction of the adjusted alpha that is optimally paid in impact depends on the concavity parameter $c$ of the price impact function. For linear price impact ($c=1$), the optimal impact is one half of the adjusted alpha, compare, e.g., \cite[p.~193]{isichenko2021}. For strictly concave price impact functions, this ratio increases and reaches two thirds for $c=1/2$ compatible with square-root impact. As a result, linear models prescribe to trade less aggressively than is optimal in their strictly concave counterparts. The impact scale $\lambda$ does not appear in the formula for the optimal impact -- it only determines how the corresponding optimal trades need to be chosen to attain the optimal impact state. 

\subsection{Implications for Trading}\label{sec:trading}

In trading applications, it is common to normalize the scaling factor $\lambda$ as 
\begin{equation}
\label{eq:lambda_def}
\lambda = \frac{\sigma}{V^c} \, g(c,\tau),
\end{equation}
where $V$ is the average daily volume ($\text{ADV}$); compare, e.g.,~\cite{Almgren2005}. This normalization expresses the quantities of interest in trader-friendly units: return predictions compare to the volatility $\sigma$ of price changes and trading volumes are expressed as fractions of the average daily volume $V$. With this normalization, the price impact coefficient also becomes comparable across different contracts; our notation $g(c,\tau)$ highlights that this estimate will depend on the corresponding impact concavity $c$ and decay time scale $\tau$.

\subsubsection*{Long-term alpha signal} We first illustrate the application of the general optimal trading result~\eqref{eq:optim} for the simplest case where $\alpha_t = \alpha$ is constant. This means that the signal predicts a return that happens in the distant future, e.g., an event at the end of the month when focusing on today's trading. For square-root impact ($c=1/2$), Equation~\eqref{eq:optim} then simplifies to
\begin{equation}\label{eq:constantalpha}
I^* = \frac{2}{3}\alpha.
\end{equation}
That is, the optimal trading strategy (for sizeable orders) pays two thirds of the constant alpha in impact. The corresponding optimal order size is
\begin{equation}\label{eq:constantalpha2}
\frac{Q_T}{V} = \Lambda^{-2} \cdot \text{sign}(\alpha) \left(\frac{\alpha}{\sigma}\right)^2, \quad \mbox{where } \quad 
\Lambda = \frac{g(1/2,\tau)}{\sqrt{1 + \frac{4}{9} \frac{T}{\tau}}}.
\end{equation}
Conversely, one implies a constant alpha from a long-term order of size $Q$ via the formula
\begin{equation}\label{eq:impliedalpha}
\frac{\alpha}{\sigma} = \Lambda \cdot \text{sign}(Q)\sqrt{\frac{|Q|}{V}}.
\end{equation}
The order size formula \ref{eq:constantalpha2} helps portfolio managers size trades considering price impact's square-root law. Then, traders use the implied alpha formula \ref{eq:impliedalpha} to agree on a baseline alpha with the portfolio team. After establishing a baseline alpha, the trading team can linearly add their own short-term signals into the trading strategy.

\subsubsection*{Mean-reverting alpha signal:} Another standard specification assumes that the alpha signal $\alpha_t$ is an Ornstein-Uhlenbeck process with relaxation time $\theta$:
\begin{equation*}
d\alpha_t  = -\frac{1}{\theta}\alpha_t dt + \sigma dW_t.
\end{equation*}
Then, $\mu^\alpha_t = -\theta^{-1}\alpha_t$, i.e., the alpha decays exponentially in the absence of new information. In this model, a statistical arbitrage portfolio continuously updates its signals based on a steady information stream. The optimal impact in this context is 
\begin{equation}\label{eq:mralpha}
I^* = \frac{2}{3}\left(1 + \frac{\tau}{\theta}\right) \alpha.
\end{equation}
Whence, mean-reverting alphas should always be traded more aggressively than constant ones, and the size of this adjustment depends on the relative magnitudes of impact decay $\tau$ and alpha decay $\theta$.

With decaying alpha, the portfolio's turnover increases compared to a constant alpha scenario, as the mean-reverting alpha case leads to higher impact states. For instance, with square-root impact, the corresponding portfolio~\eqref{eq:portfolio} trades faster over $(0,T)$ by a factor $(1 +\tau/\theta)^2$ due to the alpha's decay $\theta$.

\subsection{Implications for Transaction Cost Analysis (TCA)}

Practitioners can directly use Equation~\eqref{eq:optim} for TCA. Indeed, this optimality condition massively simplifies trade monitoring for statistical arbitrage portfolios. For instance, instead of scrutinizing each trade, portfolio managers can compare overall transaction costs with alpha signal characteristics. Verifying best execution then boils down to checking that alpha, impact, and alpha decay remain within reasonable bounds from the linear formula in Equation~\eqref{eq:optim}.

Let us illustrate with some examples how portfolio managers and traders can deploy this analysis for specific portfolios. Suppose the price impact model is of square root form ($c=1/2$) and first consider sizeable orders with constant long-term alpha. Then, in view of~\eqref{eq:constantalpha}, one  simply compares an order's transaction costs to the alpha that triggered it.

For mean-reverting alphas the optimality equation~\eqref{eq:mralpha} implies that one compares alpha capture to transaction costs considering the alpha's decay rate $\theta$ and the impact model's timescale $\tau$. The quicker the alpha's decay, the more impact the strategy has to create by trading faster in order to maximize profits.

The benefit to this impact-first approach to TCA is that, unlike in Section~\ref{sec:trading}, one does not need to share order or portfolio data to verify a trading algorithm's behavior: the entire analysis is done comparing alpha, impact, and alpha decay at the aggregate level.\footnote{\cite[Chapter~3]{webster.23} details this TCA method for linear price impact models.}

\section{Empirical Estimation}\label{sec:fitting}

\subsection{Description of the data}

The empirical analysis uses proprietary Capital Fund Management (CFM) trading data comprised of roughly $1.9 \cdot 10^5$ meta-orders of future contracts traded over 2012-2022. The time at the start and the end of each meta-order is indicated, as well as the mid-price and the number of child-orders. All meta-orders were executed through at least three child-orders and accounted for a fraction between 0.01\% and 10\% of ADV; the average order size was of the order of 0.1\%. 
No meta-order was traded longer than one day and the average execution time is 2.5h. 

\subsection{Fitting methodology}

Price returns are fitted against the increments of a power-law price impact function of the form~\eqref{eq:impact}. Since i) no meta-order in the sample lasts more than one trading day, and ii) each of them is executed with a profile close to a TWAP, the volume impact $J_t$ is computed under the assumption that trades are executed uniformly during the execution time $\Delta t$ of each meta-order, i.e.,  $Q_t = \frac{Q}{\Delta t}t$. Note that the alpha signals driving the decision are typically realized over a time window much larger than one day, implying that it is appropriate to attribute to impact, rather than alpha, the average price variation occurring during the execution of meta-orders in the sample. 

Given $J_t$ the three parameters of the AFS model are calibrated:
\begin{enumerate}[(i)]
\item The concavity coefficient $c$;
\item The impact decay timescale $\tau$;
\item The scaling factor $\lambda$, normalized as in~\eqref{eq:lambda_def}.
\end{enumerate}

\subsection{Estimating impact concavity}

\begin{figure}[th]
	\centering
	\captionsetup{width=1.0\linewidth}
    \includegraphics[width=0.5\linewidth]{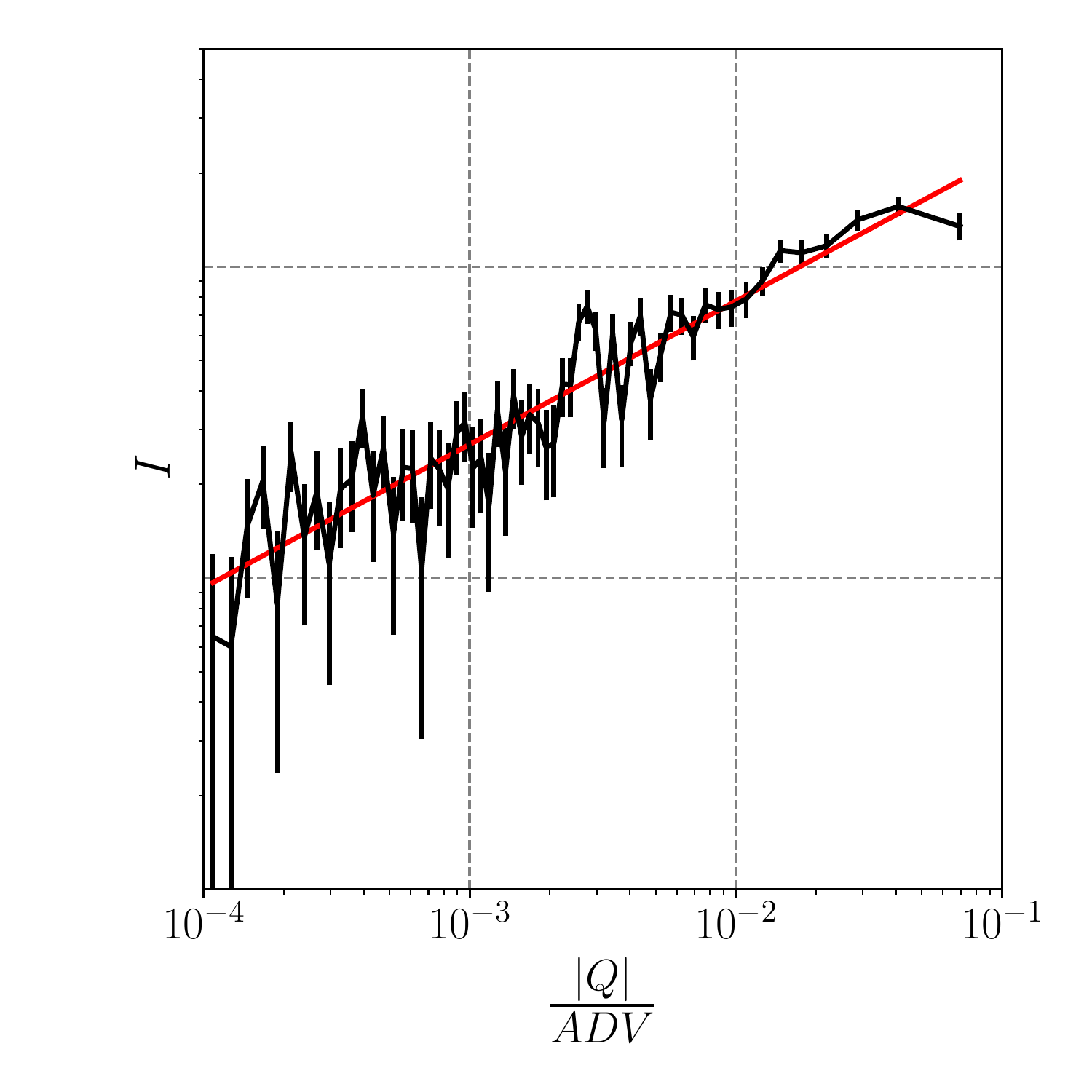}
    \caption{Expected signed return plotted against the volume fraction $|Q|/\text{ADV}$ in log-log scale. The relation is concave and follows a power-law with $c = 0.48$.}
    \label{fig:ret}
\end{figure}

As an initial visualization of price impact concavity, Figure~\ref{fig:ret} plots the average return against order size without impact decay. The log-log plot is linear, indicating a power-law relationship in the tails. The slope of the log-log fit measures the exponent of the power law for sizable orders. Here, this analysis leads to an estimate of $c = 0.48$ , consistent with the commonly observed square-root law (see e.g. \cite{BouchaudBook}). The bootstrap confidence interval from Figure \ref{fig:intro} in the introduction is computed with the standard deviation of this parameter estimate, $\textrm{std}(c)= 2.5 \cdot 10^{-2}$.

The estimation of the price impact parameters with impact decay is more subtle. To this end, we consider a grid of impact decays $\hat{\tau}$ and concavity parameters $\hat{c}$. For each pair, we then first compute the volume impact $J_t$ corresponding to the impact timescale $\hat{\tau}$ according to~\eqref{eq:volimpact}; then the price impact $I_t$ is computed for the corresponding concavity parameter $\hat{c}$. With the impact variable at hand, we then run a linear regression of price changes against impact to determine the pre-factor of the price impact model.

\begin{figure}[ht]
\captionsetup{width=1.0\linewidth}
	\center
	\begin{subfigure}[b]{0.49\textwidth}
		\includegraphics[width=1.0\linewidth]{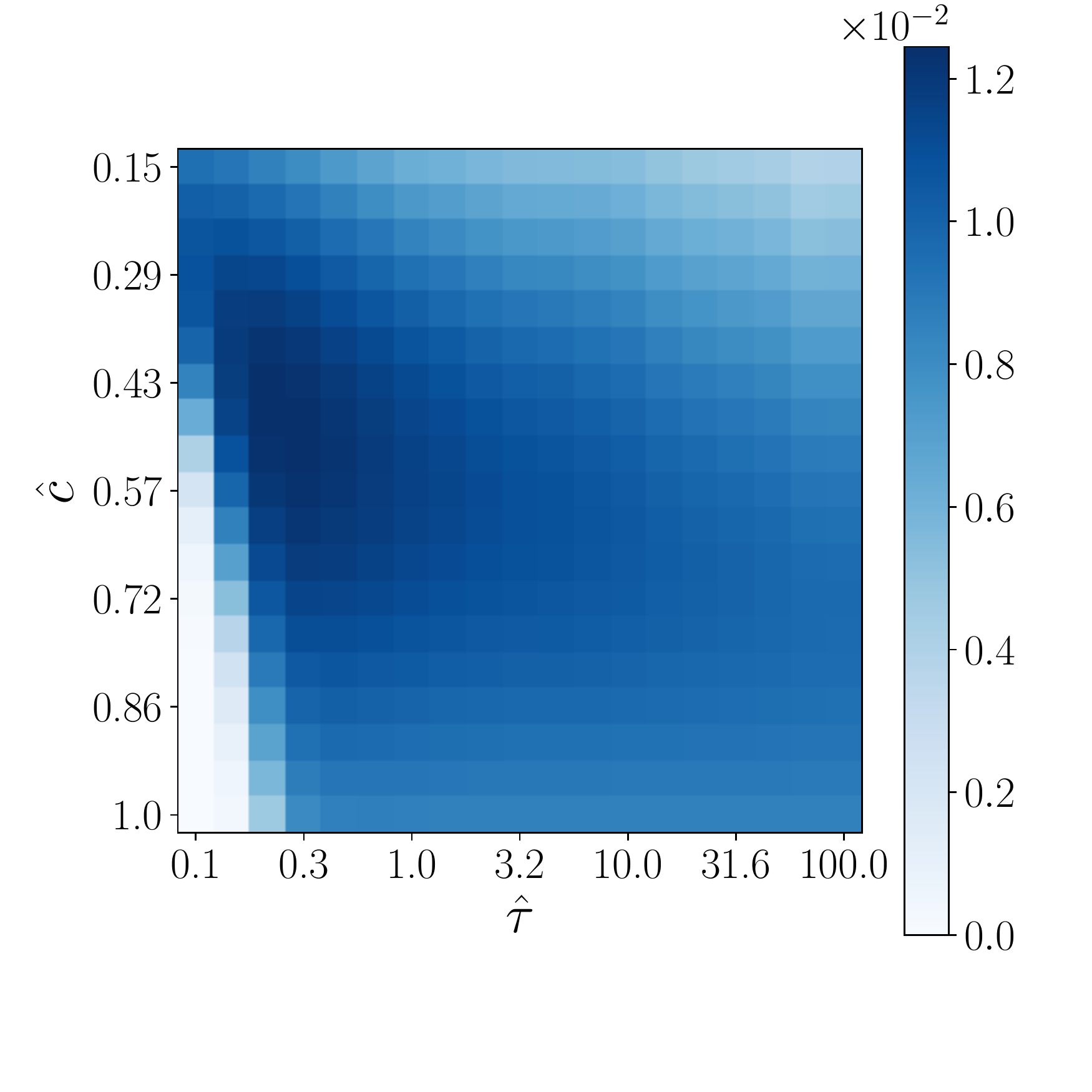}
		\caption{$R^2(\hat{c}, \hat{\tau})$.}\label{subfig:r2}
	\end{subfigure}
	\begin{subfigure}[b]{0.49\textwidth}
		\includegraphics[width=1.0\linewidth]{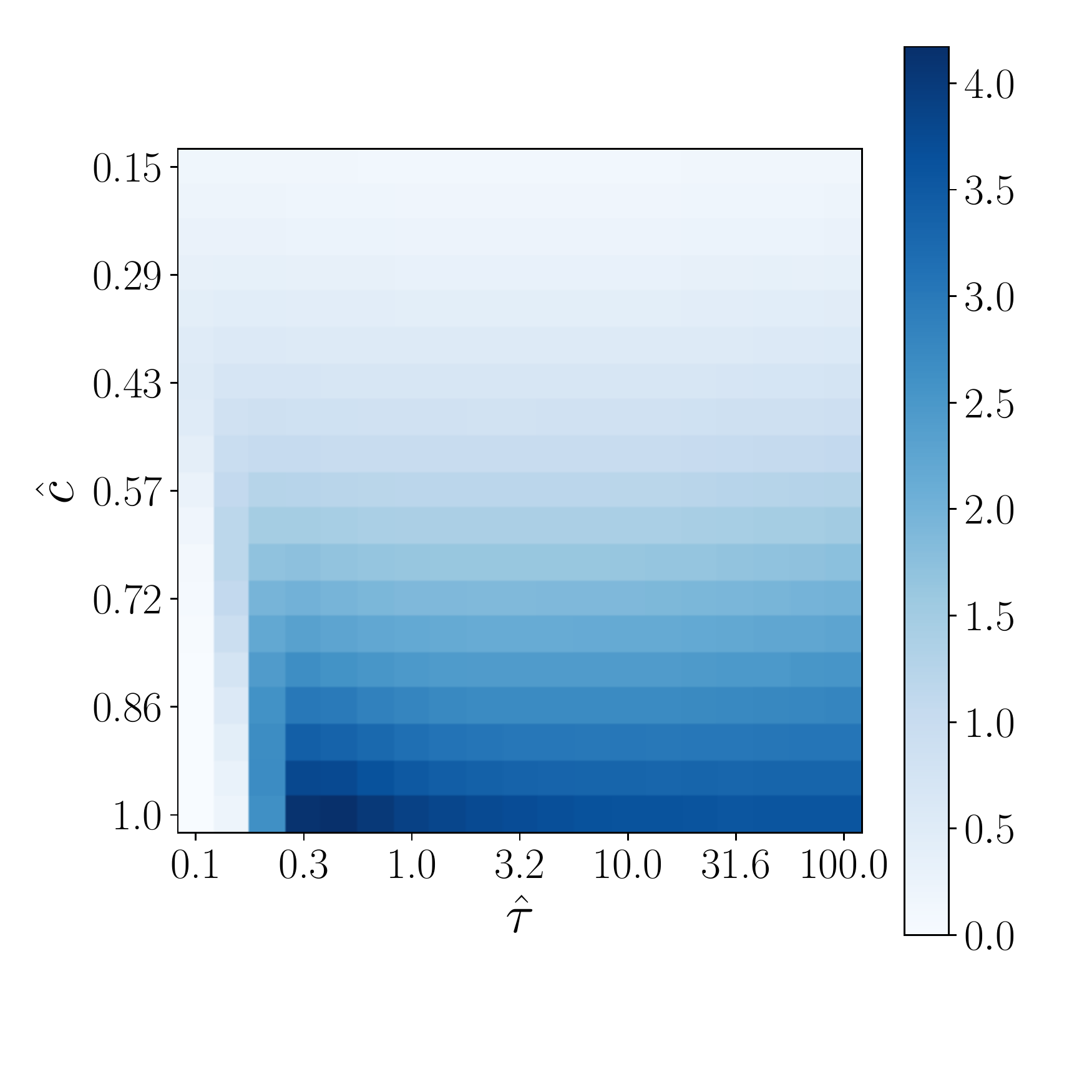}
		\caption{$g(\hat{c}, \hat{\tau})$.}\label{subfig:g}
	\end{subfigure} \\
 \begin{subfigure}[b]{0.49\textwidth}
    \includegraphics[width=1.0\linewidth]{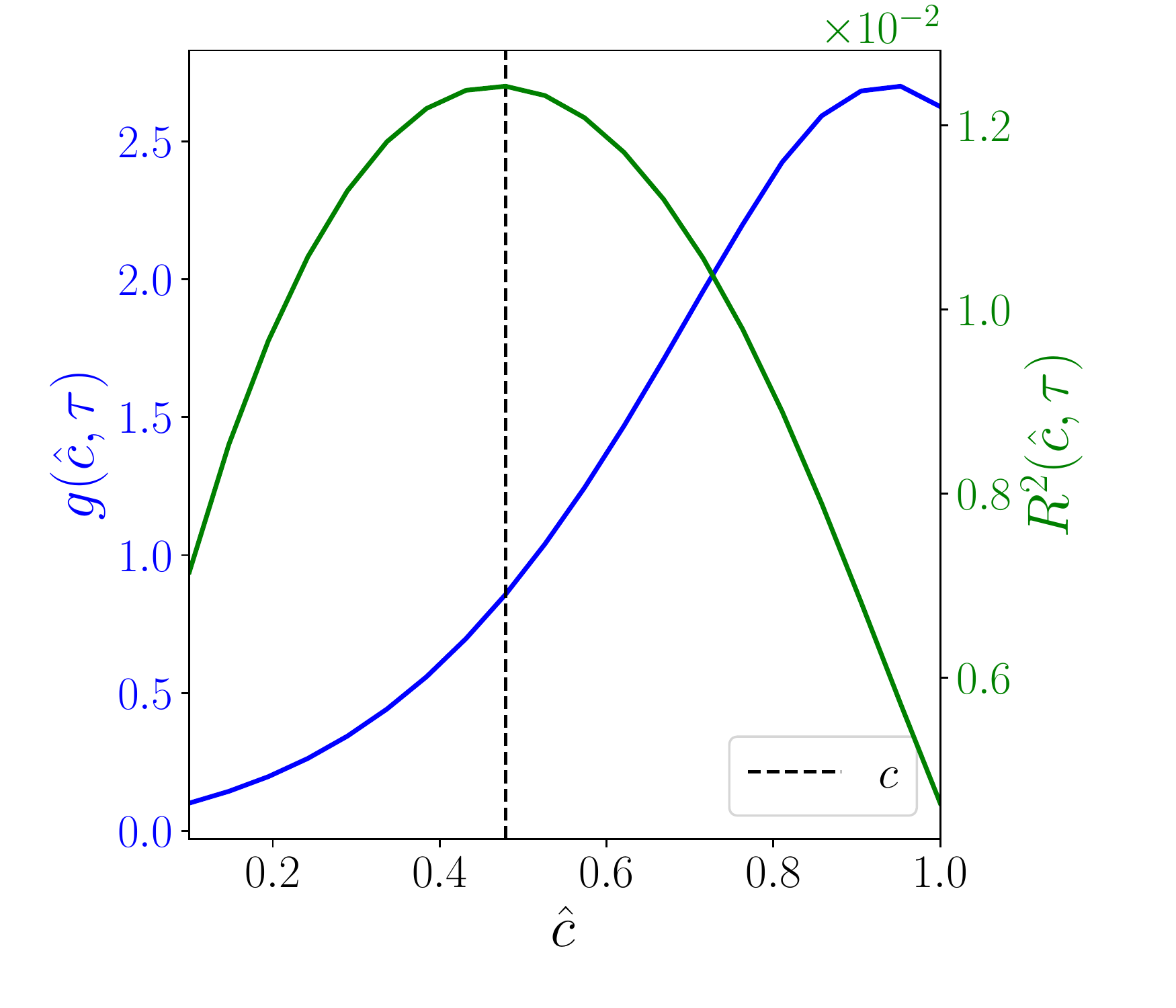}
    \caption{$R^2(\hat{c},\tau)$ and $g(\hat{c},\tau)$.}
    \label{fig:sensitivityDelta}
\end{subfigure}
	\begin{subfigure}[b]{0.49\textwidth}
		\includegraphics[width=1.0\linewidth]{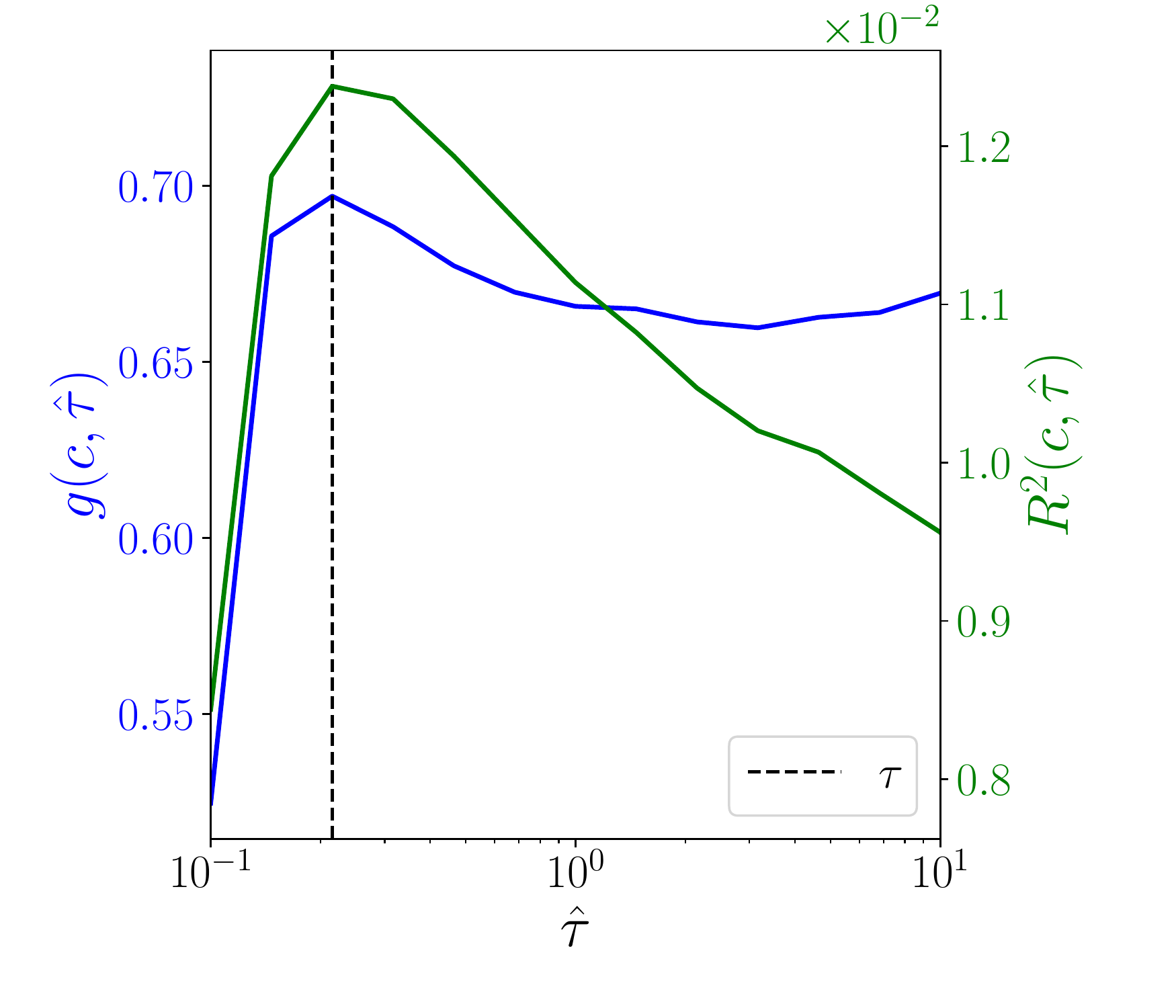}
		\caption{$R^2(c,\hat{\tau})$ and $g(c,\hat{\tau})$.}\label{subfig:tau}
	\end{subfigure}
\caption{Calibration results. Panel~(A) shows the statistical sensitivity; Panel~(B) depicts the model's pre-factor across $\hat{c}, \hat{\tau}$. The sensitivity analysis yields an ellipsoid around the best-fit parameters. On the other hand, the model's pre-factor remains roughly constant along the $\hat{\tau}$-axis and increases with  concavity $\hat{c}$. Panel~(C) shows the sensitivity across $\hat{c}$ for $\tau = 0.2$ days in green where $R^2$ peaks at the point estimate of $c = 0.48$ in line with the square-root law. The fitted pre-factor $g(\hat{c},\tau)$ is displayed in blue. Panel~(D) displays the sensitivity and the model's pre-factor across $\hat{\tau}$ for the point estimate $c$.}
    \label{fig:decay}
\end{figure}

Figure \ref{fig:decay} shows the model's statistical sensitivity $R^2(\hat{c},\hat{\tau})$ and pre-factor $g(\hat{c},\hat{\tau})$ across a broad range of concavity and decay parameters. The heatmap of $R^2$ in Subfigure \ref{subfig:r2} shows an ellipsoid around the point estimated parameters $(c,\tau)$. 
Figure \ref{fig:sensitivityDelta} plots the \emph{statistical} sensitivity of the model to the concavity parameter $c$ together with the pre-factor. 
Here, the point estimate of the decay parameter $\tau = 0.2$ days remains fixed, i.e.,  $R^2(\hat{c},\tau)$ and $g(\hat{c},\tau)$ are displayed.

As the concavity parameter $c$ varies, the model's $R^2$ peaks at $c = 0.48$ and decreases markedly around this point estimate.\footnote{The magnitude of the model's $R^2$ is consistent with results published in papers using the public trading tape. Indeed, when one runs the price impact model on the full trading tape, the $R^2$ varies between 10\% and 20\%. However, in this study, only CFM's metaorders are used. Therefore, the $R^2$ drops in line with the fraction of the public tape CFM's orders represent.} The pre-factor $g(\hat{c},\tau)$ increases with increasing concavity parameter. This increase is because the brunt of the data consists of smaller orders, leading to the linear model under-estimating impact compared to a concave model.

\subsection{Estimating impact decay}
To investigate the statistical sensitivity with respect to the impact decay, the concavity parameter $c$ is fixed at the point estimate $c=0.48$ and the impact decay $\hat{\tau}$ varies. Figure \ref{subfig:tau} displays $R^2(c,\hat{\tau})$ and $g(c,\hat{\tau})$. There is a peak at $\tau = 0.2$ days, indicated that one measures the short-time scale of impact decay. The model's pre-factor remains relatively constant for timescales larger than 0.1 days.

The primary take-away is that the \emph{statistical} loss is significantly more sensitive to impact concavity than decay. For example, misspecifying $c = 1$ instead of the point estimate $c=0.48$ reduces $R^2$ by 50\%. Conversely, getting $\tau$ wrong by a factor of ten reduces $R^2$ by less than 2\%.\footnote{This is not a new observation: studies across many datasets and asset classes agree on price impact's concavity and general order of magnitude but disagree on decay. The reason is that different orders decay at different timescales, which are better captured by a multi-exponential or power law decay kernel. With linear price impact, there are some results for such decay kernels, cf.~\cite{Bouchaud2004,gatheral.al.12,abi.neuman.22} and the references therein. Extending this analysis to models with nonlinear impact \emph{and} impact decay is an important direction for future research}

These empirical results do \emph{not} imply that calibrating impact decay correctly is unimportant. Indeed, a central result in Section \ref{sec:sensitivity} is that a portfolio's P\&L can be \emph{highly} sensitive to $\tau$, even if \emph{statistically}, $R^2$ is not as sensitive to $\tau$.

\section{Sensitivity Analysis}\label{sec:sensitivity}

We now combine the theoretical and empirical results from the previous sections to quantify the performance losses caused by trading based on a misspecified model.

\subsection{Quantifying misspecification costs}

Recall that, for given estimates $\hat{c}, \hat{\tau}$ of impact concavity and decay, the corresponding fitted pre-factor is 
\begin{equation*}
\lambda(\hat{c}, \hat{\tau})  = \frac{\sigma}{V^{\hat{c}}} g(\hat{c}, \hat\tau).
\end{equation*}
Therefore, assuming $\hat{c}, \hat\tau$ are the correct model parameters, a portfolio trading an alpha signal $\alpha_t$ with decay $\mu^\alpha_t$ will implement
\begin{equation*}
I_t(\hat{c}, \hat\tau) = \frac{\alpha_t - \hat\tau \mu^\alpha_t}{1+\hat{c}} \quad \mbox{or, equivalently,} \quad J_t(\hat{c}, \hat\tau) = \left(\frac{\alpha_t - \hat\tau \mu^\alpha_t}{(1+\hat{c})\lambda(\hat{c}, \hat\tau)}\right)^{1/\hat{c}}.
\end{equation*}
We call $I(\hat{c}, \hat\tau)$ or, equivalently, $J(\hat{c}, \hat\tau)$, the \emph{misspecified} policy. This is the optimal policy when the impact parameters $\hat{c}, \hat\tau$ correctly describe price impact, but not if these parameters differ from the price impact parameters $c, \tau$ of the \emph{actual} data generating model. To ease notation, we omit the corresponding variables when one of the parameters is held fixed at its actual value, e.g., we use the shorthand $I(\hat\tau)=I(c, \hat \tau)$.

The P\&L of the \emph{misspecified} policy $J(\hat{c},\hat\tau)$ under the \emph{actual} price impact model with parameters $c$, $\tau$ is 
\begin{align*}
U\Big(J(\hat{c}, \hat \tau); c, \tau\Big)= &\frac{1}{\lambda}\E\Bigg[ \frac{1}{\tau}\int_0^T ((\alpha_t - \tau \mu^\alpha_t) J_t(\hat{c}, \hat\tau) - \lambda |J_t(\hat{c}, \hat\tau)|^{1+c}) dt\\
&\qquad + \alpha_T J_T(\hat{c}, \hat\tau) - \frac{1}{1+c}|J_T(\hat{c}, \hat\tau)|^{1+c}\Bigg].
\end{align*}
This general formula is straightforward to implement numerically in a back test. For instance, statistical arbitrage teams can plug in historical alpha signals to measure the expected P\&L of trading based on $\hat{c}, \hat \tau$ when the actual parameters are $c, \tau$.

To illustrate the misspecification formula's implications, Sections \ref{sec:missConcavity} and \ref{sec:missDecay} cover two common cases. By specifying concrete parametric alpha signals, the cost of misspecifying concavity or decay can be quantified in closed-form, which reveals the dependence on characteristics of the corresponding alpha signals without relying on a back test.

\subsection{Misspecification cost of impact concavity}\label{sec:missConcavity}

We first consider the simplest case of a constant alpha signal: $\alpha_t = \alpha$, so that $\mu^\alpha_t = 0$. For example, this assumption is reasonable when trading intraday based on a signal that predicts a return past today's close, such as an event taking place the next day or week. 

As impact decay only plays a minor role in this context,\footnote{Indeed, for constant alpha, the P\&L of the believed strategy only depends on the impact decay parameter $\hat{\tau}$ through the estimate of the magnitude of price impact $\lambda(\hat{c},\hat{\tau})$. As this dependence is rather weak, impact decay only plays a minor role in the absence of alpha decay.}  we assume for simplicity that it is correctly specified ($\hat \tau = \tau$) and focus on the misspecification cost of the concavity parameter $\hat c \neq c$. In words: 
\begin{displayquote}
How much P\&L does the portfolio lose if it trades with the wrong price impact concavity?
\end{displayquote}
For example, one canonical case is when $c = 0.5$ and $\hat c = 1$: then, one estimates the cost of implementing a linear price impact model when actual impact follows a square-root law.

The expected P\&L for the \emph{misspecified} policy is
\begin{align*}
U\Big(J(\hat{c}); c\Big)=  & \frac{\sigma \,V}{g(\hat{c})^{1/\hat{c}}} \Bigg[\left(\frac{\alpha}{\sigma}\right)^{1+1/\hat{c}} \left(\frac{T}{\tau (1 + \hat{c})^{1/\hat{c}}} + 1\right)\\
&\qquad\qquad - \frac{g(c)}{g(\hat{c})^{c/\hat{c}}} \left(\frac{\alpha}{\sigma}\right)^{(1+c)/\hat{c}} \left(\frac{T}{\tau (1+\hat{c})^{(1+c)/\hat{c}}} + \frac{1}{1+c}\right) \Bigg].
\end{align*} 

\begin{figure}[ht]
    \includegraphics[width=0.9\linewidth]{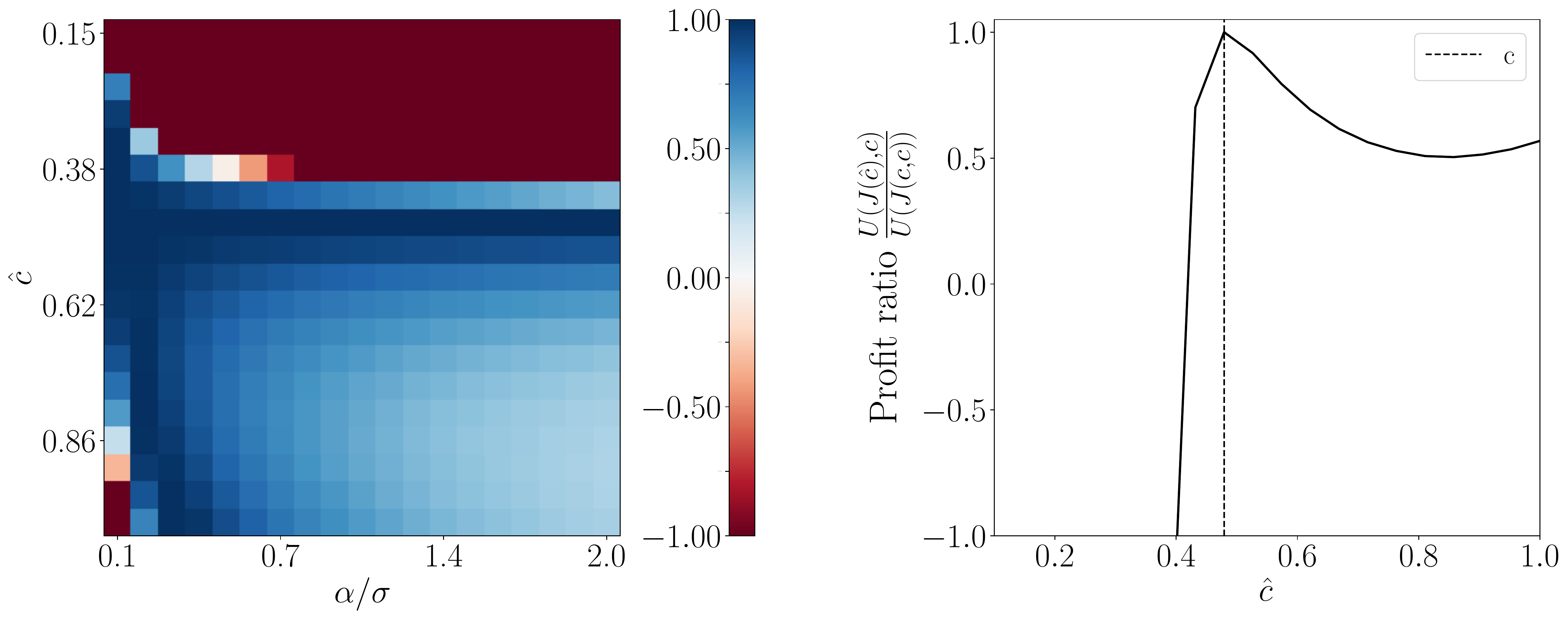}
\caption{Profit ratios between correctly and incorrectly specified strategies when $c = 0.48$. Left panel: Heat map of the profit ratio across $\hat c$ and $\alpha/\sigma$. Right panel: Profit ratio across $\hat{c}$ for an alpha signal with Sharpe $\alpha/\sigma = 1$. The profits are asymmetric: over-estimating concavity (low $\hat\delta$) quickly leads to losses, while under-estimating concavity only shrinks profits. } 
    \label{fig:misspecConcavity}
\end{figure}

The natural comparison for this is the expected P\&L of the optimal policy for the actual concavity parameter $c$:
\begin{equation*}
U\Big(J(c); c\Big) = \frac{\sigma \,V}{g(c)^{1/c}} \left(\frac{\alpha}{\sigma}\right)^{1+1/c} \frac{c}{1+c} \left(\frac{T}{\tau (1+c)^{1/c}} + 1\right). 
\end{equation*}
In this setting, the primary alpha characteristic is the signal's Sharpe ratio, $\alpha/\sigma$. A core result from Figure~\ref{fig:misspecConcavity} is that misspecification costs are more sensitive to $c$ as a signal's Sharpe ratio increases. Therefore, the stronger a team's alpha signal, the more important it becomes to correctly estimate the price impact model's concavity accurately. 

In addition to quantifying the empirical sensitivity of P\&L to misspecifications of the concavity parameter $c$, the closed-form formulas also allow one to compute \emph{how wrong} a parameter can become before turning profitable strategies unprofitable. Indeed, for each true parameter $c$, there exists a critical value $c_{\min}$ such that, for any choice  $\hat c < c_{\min}$, the expected P\&L becomes a \emph{decreasing} function of the alpha signal. Naturally, statistical arbitrage portfolios crucially need to avoid this critical regime where adding more predictive power leads to worse outcomes. Figure \ref{fig:misspecConcavity} illustrates the sharpness of the phase transition between profitable and non-profitable misspecified trading strategies. Because misspecification costs become more sensitive with higher alpha Sharpe ratios, the critical value becomes tighter as the Sharpe increases. Again, this highlights the importance of correctly estimating price impact's concavity when trading high Sharpe ratio signals.

A crucial insight is that this misspecification risk is asymmetric: even for large over-estimates $\hat c > c$, the expected P\&L typically remains positive (except for very small Sharpe ratios). The intuition is that overestimating the concavity parameter $c$ leads the portfolio to submit smaller trades. While this is suboptimal and sacrifices some profit opportunities, it will only lead to shrinking profits but not losses. Conversely, underestimates below the critical value $c_{\min} < c$ can lead to significant losses, as the corresponding trading strategy submits outsized trades leading to excessive trading costs. 

While these results are derived in a concrete model with a convenient closed-form solution, we expect them to remain true qualitatively across a much wider range of models. This leads to a practical takeaway when fitting price impact models: when considering confidence intervals for a price impact model's concavity, the robust solution is to deploy a more conservative estimate from the upper part of the confidence interval, i.e., use a impact model that is somewhat less concave than the point estimate. 

\subsection{Misspecification cost of impact decay}\label{sec:missDecay}

We now turn to misspecification of the impact decay parameter $\tau$, while keeping the concavity parameter $c$ fixed to its correct value. As impact decay predominantly interacts with the alpha decay, we study this for a mean-reverting alpha with $\mu^\alpha_t = -\alpha_t/\theta$. To obtain crisp results, we focus on the steady-state limit $\lim_{T\to \infty} \frac1T \mathbb{E}[Y_T]$ of the P\&L. Unlike the model's statistical $R^2$, it turns out that the expected P\&L is surprisingly sensitive to misspecifications in $\tau$.

For the \emph{misspecified} policy, we have
\begin{align*}
U\Big(J(\hat\tau); \tau\Big) = &\frac{\sigma \,V}{\tau}\left(\frac{1 +\hat\tau/\theta}{g(\hat\tau)(1+c)}\right)^{1/c}\left(1 + \frac{\tau}{\theta} - \frac{g(\tau)(1 +\hat\tau/\theta)}{g(\hat\tau)(1+c)}\right)\\
&\qquad \times\lim_{T\to \infty} \frac{1}{T}\left[\int_0^T \left|\frac{\alpha_t}{\sigma}\right|^{1+1/c} dt \right].
\end{align*}
The optimal value derived from the policy for the actual impact decay $\tau$ is 
\begin{align*}
U\Big(J(\tau); \tau\Big) = &\frac{\sigma \,V}{\tau} \frac{c(1+\tau/\theta )^{1+1/c}}{g(\tau)^{1/c} (1+c)^{1+1/c}}\times\lim_{T\to \infty}\frac{1}{T}\left[\int_0^T \left|\frac{\alpha_t}{\sigma}\right|^{1+1/c} dt \right].
\end{align*}

In this setting, the key alpha characteristic is its decay $\theta$. More specifically, the ratio of performances for the misspecified and optimal policies mostly depends on the estimated and true impact decays through their ratios relative to alpha decay, $\hat \tau/\theta$ and $\tau/\theta$. Indeed, as depicted in Figure~\ref{subfig:tau}, $g(\tau)\approx g(\hat{\tau})$ is a good approximation for a wide range of decay parameters; with this, the ratio of performance simplifies to 

\begin{align*}
  \frac{U\Big(J(\hat \tau); \tau\Big)}{U\Big(J(\tau); \tau\Big)} = \frac{1}{c} \left(\frac{1+\hat \tau / \theta}{1 + \tau/\theta }\right)^{1/c} \left(1+c -  \frac{1+\hat \tau / \theta}{1 + \tau/\theta} \right).
\end{align*}

\begin{figure}[ht]
    \includegraphics[width=0.9\linewidth]{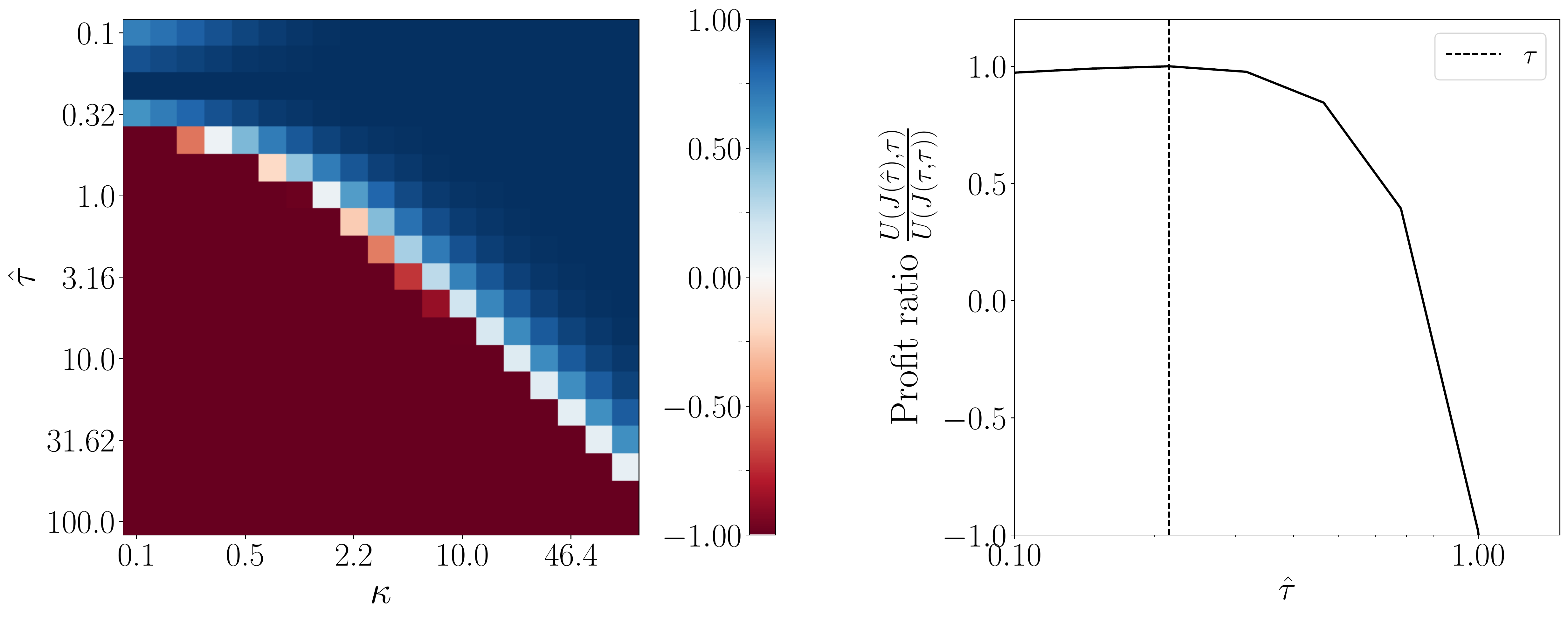}
\caption{Profit ratio between correctly and incorrectly specified strategies when $\tau = 0.2$ days. Left panel: Heat map of the profit ratio across $\hat \tau$ and $\theta$ in log-log scale. Right panel: Profit ratio across $\hat \tau$ in log-scale for an alpha signal with a decay timescale of 1 day.  The profits are asymmetric: over-estimating decay (high $\hat\tau$) quickly leads to losses, while under-estimating decay only shrinks profits.} 
    \label{fig:misspecDecay}
\end{figure}

The profit ratio between misspecified and optimal policy is displayed in Figure~\ref{fig:misspecDecay}. There is a sharp boundary between the profitable and unprofitable regions that depends linearly on the ratios $\hat \tau/\theta$ and $\tau/\theta$. This boundary appears then curved in log-log scale on the left panel in Figure \ref{fig:misspecDecay} separating the positive and negative profit regions. A wrong estimation of impact decay for a long-lasting alpha signal (small $\theta$) is less costly than for a fast decaying one. 

Like for impact concavity, the costs of misspecifying impact decay are highly asymmetric. Indeed, overestimating impact decay quickly turns profitable alpha signals unprofitable, as illustrated in the right panel of Figure~\ref{fig:misspecDecay} for a moderate alpha decay time scale of 1 day. The intuition is that overestimated $\hat{\tau}$ lead to larger than optimal impacts, i.e., overly aggressive trading just like underestimates of the concavity parameter. 

\section{Conclusion}
In this paper, we investigate the effect that the misspecification of the price impact model has on the performance of a trader. This leads to three broad takeaways: 
\begin{enumerate}[(i)]
\item For the AFS model, the closed-form formula
\begin{equation*}
I^*_t = \frac{1}{1+c}\left(\alpha_t - \tau \mu^\alpha_t\right)
\end{equation*}
relates the optimal impact state $I^*$ to a trader's alpha level $\alpha_t$, alpha decay $\mu^\alpha_t$, impact concavity $c$, and impact decay $\tau$.
\item Statistical estimates yield $c \approx 0.5$ for sizeable meta-orders, in consensus with the literature~\citep{Almgren2005,BouchaudBook}. However, statistical estimates of $\tau$ are less precise: a broad range of impact decays $\tau$ statistically fit the model similarly well. 
\item The P\&L cost of getting an impact parameter wrong can be evaluated using the AFS model's tractable \emph{misspecification cost} formula. This P\&L-driven approach complements a statistical approach and shows that the opportunity costs of getting model parameters wrong are \emph{asymmetric}:
\begin{itemize}
\item It is better to over- than under-estimate $c$. An excessively concave price impact model can lead to losses.
\item It is better to under- than over-estimate $\tau$. A price impact model with excessively fast impact decay can lead to losses.
\end{itemize}
In both cases, overly aggressive trading has a substantially bigger effect than an overly cautious approach. While we illustrate this with the closed-form solutions for the AFS model, these insights are expected to remain qualitatively true for many models. Therefore, a general approach when considering confidence intervals for price impact model parameters is to deploy parameters within the error band that lead to more conservative trades.
\end{enumerate} 

Our results thereby support the relevance of a robust approach to optimal trading problems in the presence of trading costs, and supports the idea of jointly tackling parameter estimation and alpha exploitation.

\maketitle
\bibliographystyle{plainnat}
\bibliography{biblio}

\end{document}